\documentstyle[12pt,epsfig]{article}
\let\section=\subsection     \let\subsection=\subsubsection                

\begin{document}
\begin{center}
   {\large \bf CAN EXTENDED DUALITY UNIQUELY EXPLAIN}\\[2mm]
   {\large \bf THE DILEPTON DATA IN HIC'S AT SPS?}\\[5mm]
   B.~K\"AMPFER, K.~GALLMEISTER, O.P.~PAVLENKO \\[5mm]
   {\small \it  Forschungszentrum Rossendorf\\
   Postfach 510119, 01314 Dresden, Germany \\[8mm] }
\end{center}

\begin{abstract}\noindent
Extended duality 
is used to explain the dilepton data of heavy-ion experiments at CERN-SPS.
Taking into account the background contributions (hadronic cocktail,
Drell-Yan, semileptonic decays of open charm) the spectral shapes of the
CERES, NA38/50 and HELIOS-3 data in experiments with lead and sulfur beams
can be well described, however, the normalizations of the sulfur beam data
are problematic.
\end{abstract}

\section{Introduction}
The typical temperature scales of hadron matter in heavy-ion collisions
at CERN-SPS energies, extracted from hadron abundances and hadron transverse
momentum spectra, are in the order of 
$\langle T \rangle \approx T_c \sim {\cal O} (m_\pi)$, where
$T_c$ stands for the expected deconfinement and/or chiral symmetry
restoration temperature. It is the chiral symmetry restoration which is
thought to cause a drastic reshaping of the low-lying hadron spectrum:
In a low-temperature expansion the vector and axial-vector parts of
the current-current correlator, which determines the emissivity of matter,
are mixed and cause a dilepton spectrum which resembles a simple
$q \bar q$ rate at given temperature for invariant masses
$M > 1$ GeV (cf.\ fig.~2.9 in \cite{RappWambach});
at lower invariant masses the various interaction processes in hot matter
cause a disappearance of pronounced structures emanating from the lowest
vector mesons, and also here the resulting spectrum can be approximated
by the $q \bar q$ rate (cf.\ fig.~4.6 in \cite{RappWambach}).
As a result, the descriptions of a dense hadron system and a quark-gluon
system become dual to each other.
 
Here we employ this duality and analyze the dilepton spectra
obtained in sulfur and lead beam reactions at the CERN-SPS.
Since we use duality (i) in the full invariant mass range considered and
(ii) for the full time evolution of matter we phrase our approach as
a test of the extended duality hypothesis. 

\section{The model}

Since the experiments have no space-time resolution we parameterize
the dilepton spectra by
\begin{equation}
\frac{dN}{p_{\perp 1} \, d p_{\perp 1} \, 
p_{\perp 2} \, d p_{\perp 2} dy_1 \, dy_2 \,
d \phi_1 \, d \phi_2}
= 
\frac{5 \alpha^2}{72 \pi^5} 
N_{\rm eff} 
\exp \left\{ - \frac{M_\perp \cosh (Y - Y_{cms})}{T_{\rm eff}} \right\},
\end{equation}
where $p_{\perp 1,2}$, $y_{1,2}$ and $\phi_{1,2}$ denote the transverse
momenta, rapidities and azimuthal angles of the individual leptons 1 and 2,
which must be appropriately combined to construct the pair mass $M$,
the pair transverse momentum $Q_\perp$ and transverse mass $M_\perp$.
This expression stems from the $q \bar q$-annihilation
dilepton yield of a source localized
at midrapidity $Y_{cms}$. The two parameters $T_{\rm eff}$ and
$N_{\rm eff}$ are to be adjusted to the experiment.
In a more detailed description one has to introduce the space-time
dependence of the source parameters such as volume, temperature,
chemical potentials and matter velocity
(cf.\ \cite{KlinglWeise,RappWambach_new,RappShuryak,Phys.Lett.})
and to integrate over the evolution up to freeze-out. In this way the 
two parameters $T_{\rm eff}$ and $N_{\rm eff}$ are mapped on a much larger
parameter space, which however is constrained by hadronic observables
\cite{RappWambach_new,RappShuryak} and allows a detailed explanation of 
$T_{\rm eff}$ and $N_{\rm eff}$. According to our experience the 
detailed space-time models do not improve the description
of the data discussed below (cf.\ \cite{RappWambach_new,RappShuryak}
vs.\ \cite{Phys.Lett.}). As matter of fact we mention that the
dilepton rate, in particular the transverse momentum spectrum, 
depends sensitively on the flow of matter when
using the correct Lorentz invariant expression \cite{PRC_95,Phys.Lett.}.
Fortunately, the transverse flow is constrained by hadron data
\cite{J.Phys.,Heinz}, and for realistic scenarios the flow is masked by
background dilepton contributions.

\section{Analysis of dilepton spectra}

\subsection{Lead beam data}

A comparison of the model (1) with lead beam data is displayed in
fig.~1. With a uniquely fixed set of $T_{\rm eff} =$ 170 MeV and 
$N_{\rm eff} = 3.3 \times 10^4$ fm${}^4$ a fairly well description of the data
is accomplished. We mention that the
use of the hadronic decay cocktail 
(dashed curves in the upper panels in fig.~1)
and the normalization of
$\frac{d N_{ch}}{d \eta} = 250$ are essential for describing the CERES
$e^+ e^-$ data \cite{CERES_Pb} in central reactions Pb(158 AGeV) + Au.
For a description of the NA50 $\mu^+ \mu^-$ data \cite{NA50} 
in central reactions Pb(158 AGeV) + Pb, the Drell-Yan contribution 
(dashed curves in the lower panels in fig.~1)
and the correlated semileptonic decays of open charm mesons 
(dot-dashed curves in the lower panels in fig.~1)
are needed.
The latter ones are generated with PYTHIA \cite{PYTHIA}
(D-' structure functions,
charm mass 1.5 GeV, default fragmentation ''hybrid''), 
where the Drell-Yan
K factor of 1.2 is adjusted to the data \cite{DY_K_factor} and
the open charm K factor of 4 to the compilation of identified hadronic
charm channels in \cite{PBM}. Confidence in these important K factors
is gained by a comparison with $\mu^+ \mu^-$ data in the reaction
p(450 GeV) + W \cite{pW_Capelli}. Note that we here exploit the approximate
NA50 acceptance according to \cite{RappShuryak} for the lead beam data.
To translate the cross sections delivered by PYTHIA 
into rates we use a thickness function of 31 mb${}^{-1}$
for central collisions Pb + Pb.
Also $J/\psi$ and $\psi '$ contributions \`a la \cite{NA50} are included.

\begin{center}
\centering
~\\[-.6cm]
\hspace*{6mm}
\psfig{file=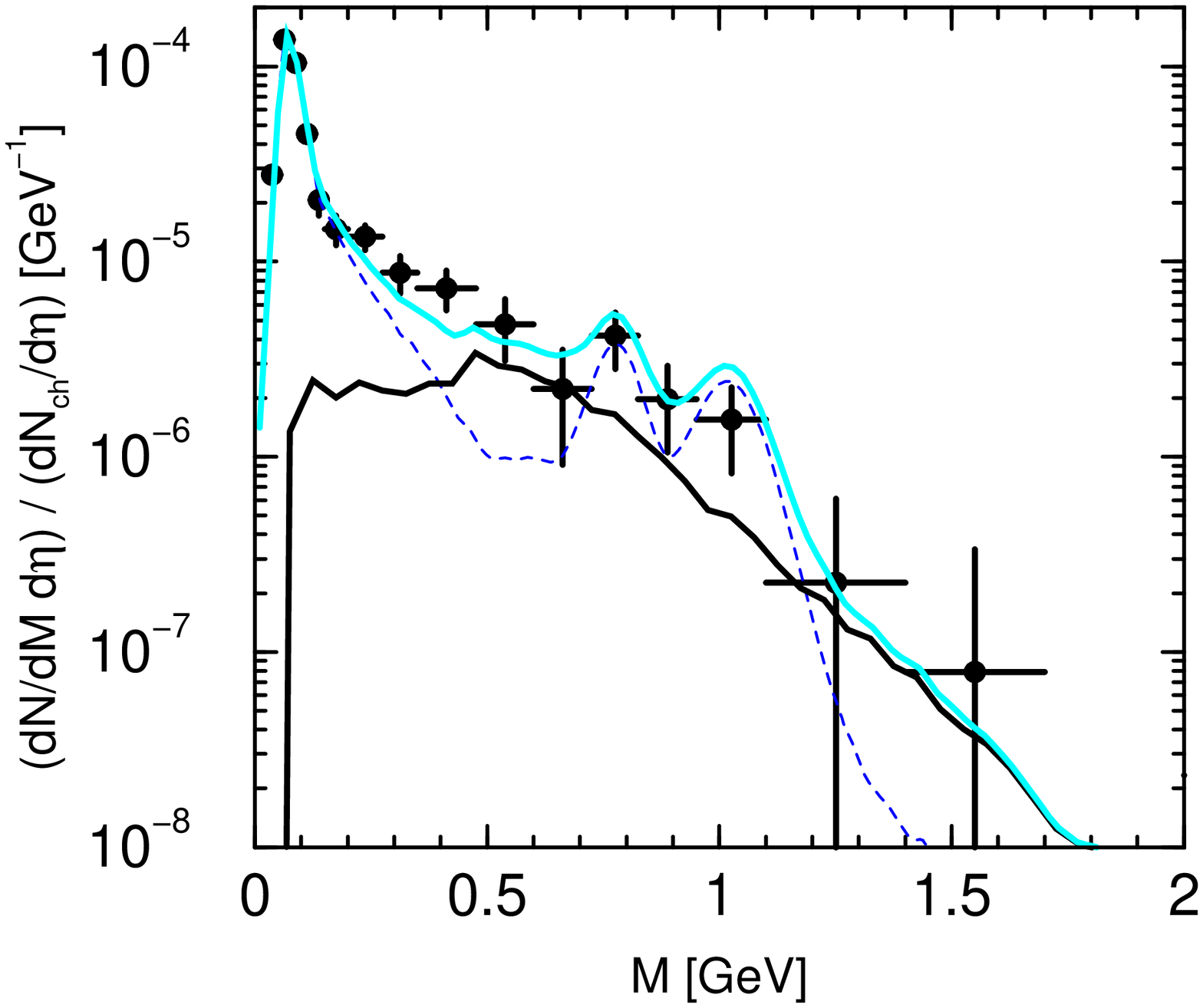,width=5.3cm,angle=-90}
\hfill
\psfig{file=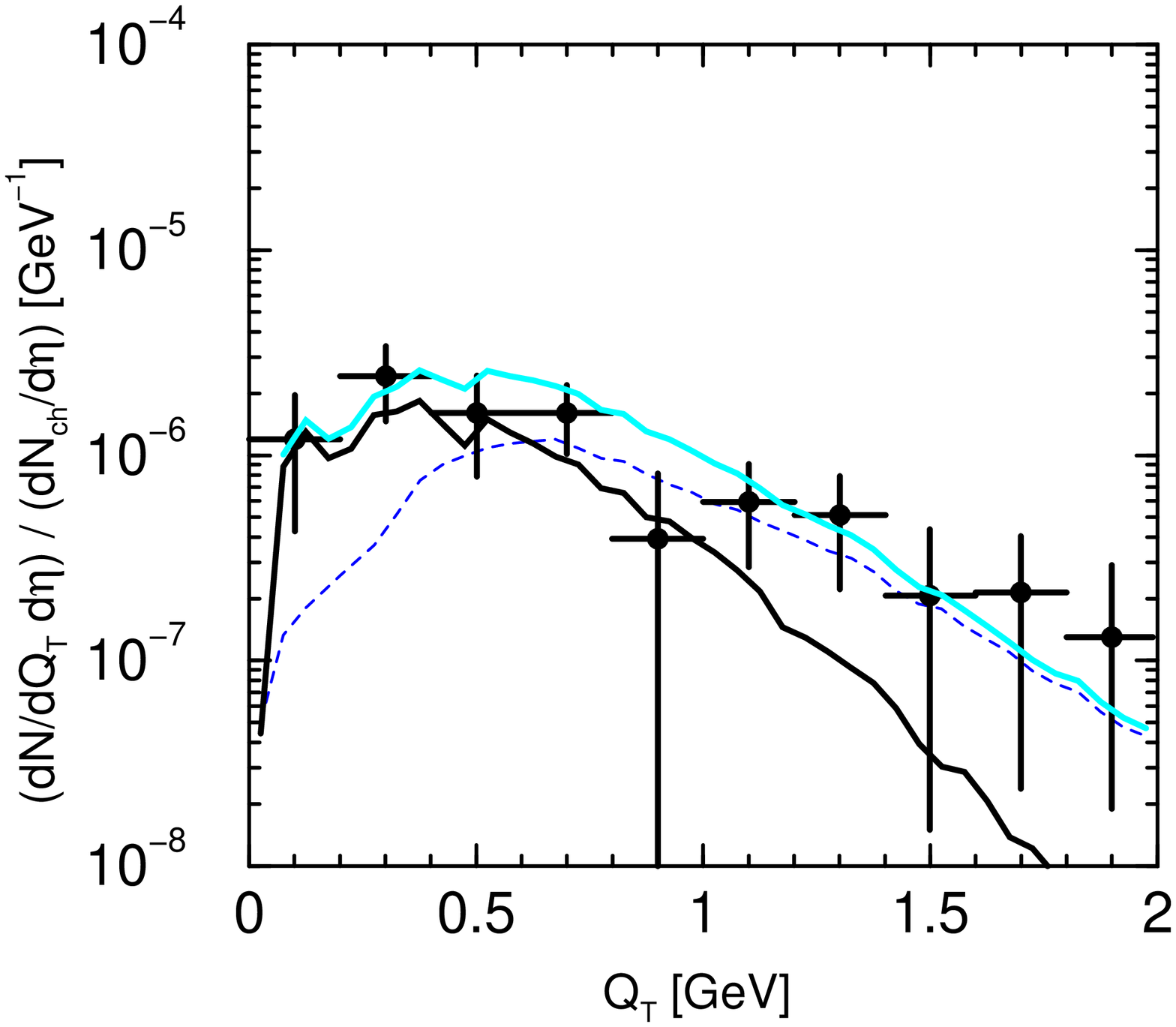,width=5.3cm,angle=-90}
\hspace*{6mm}

\hspace*{6mm}
\psfig{file=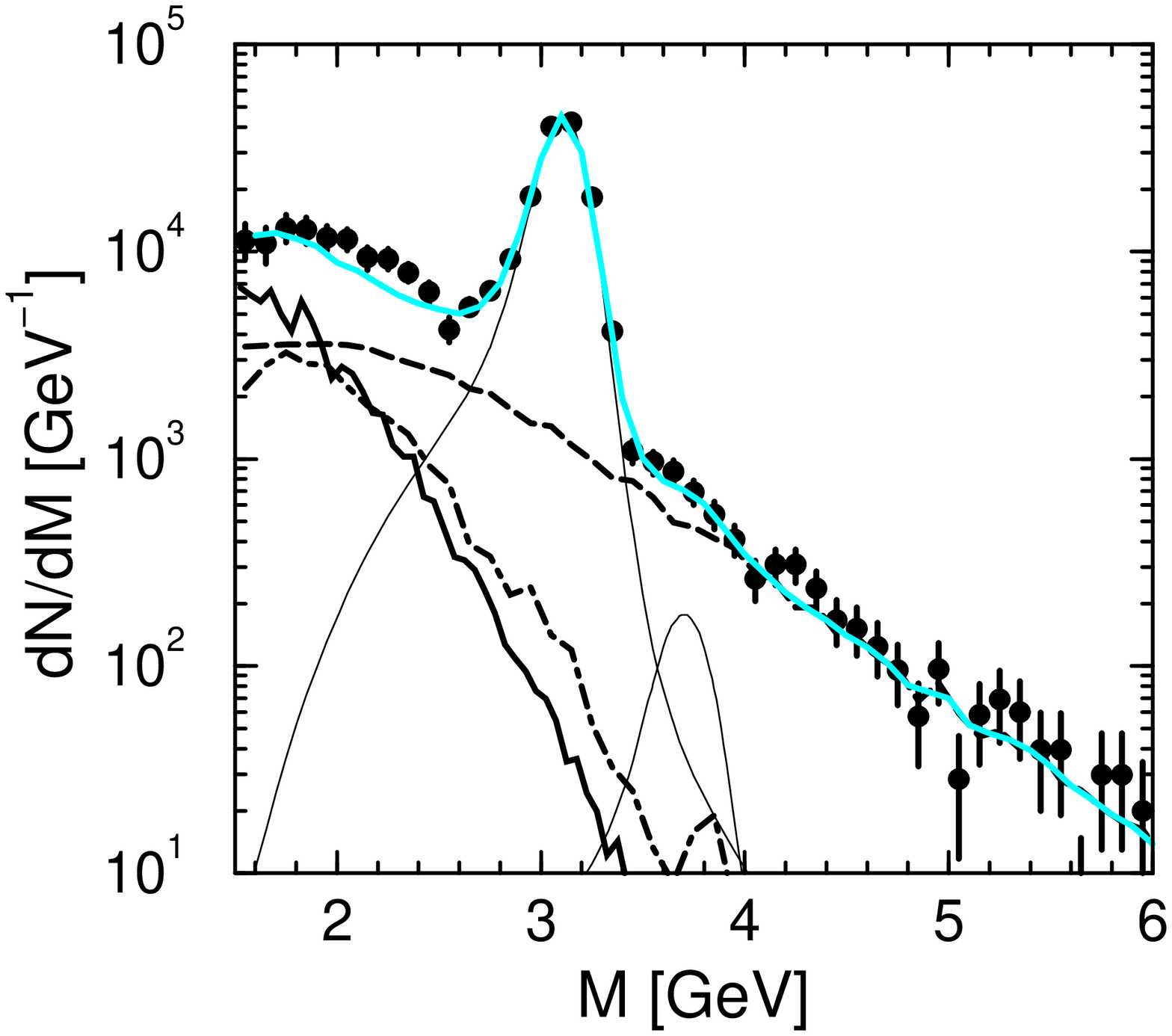,width=5.3cm,angle=-90}
\hfill
\psfig{file=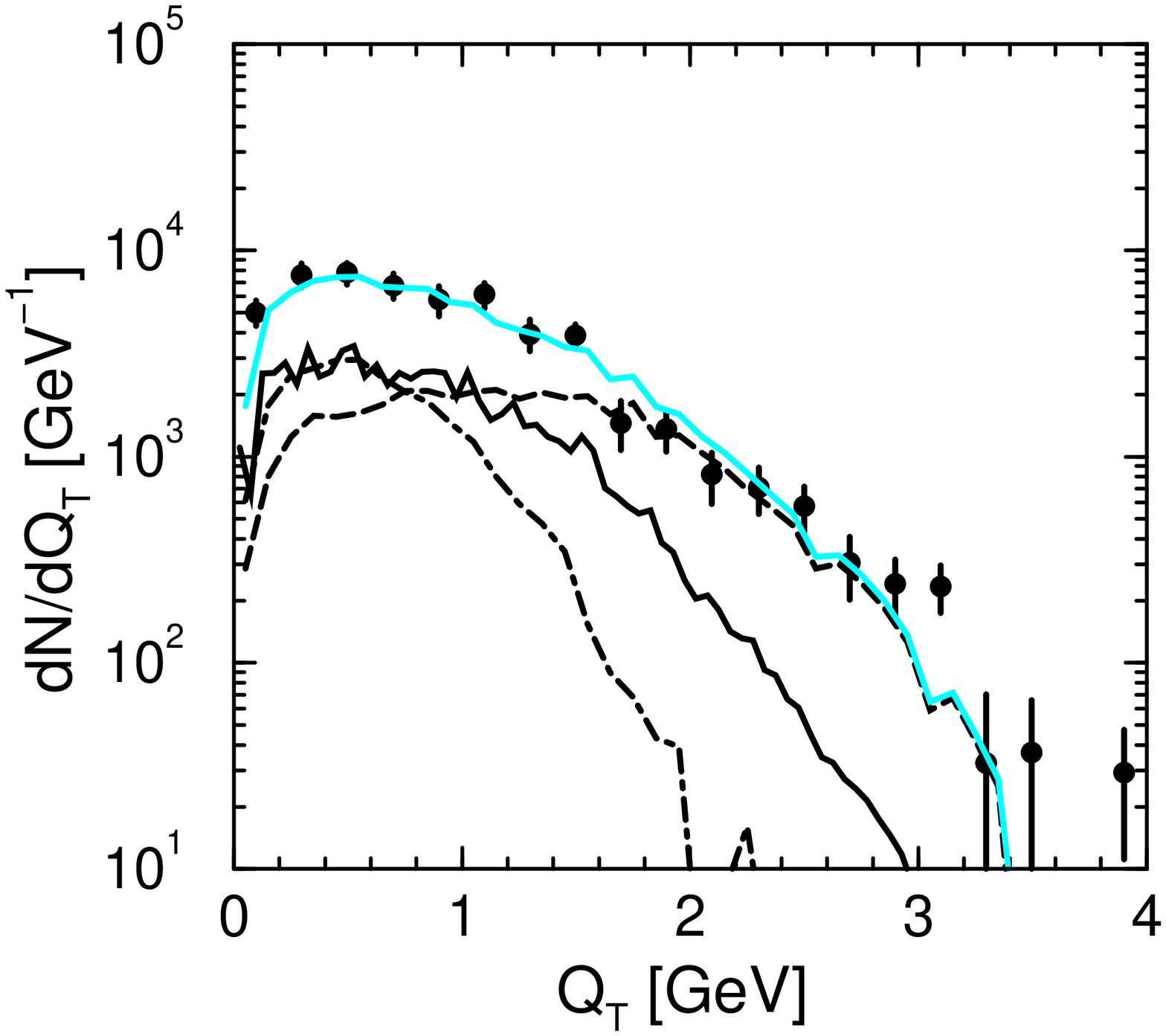,width=5.3cm,angle=-90}
\hspace*{6mm}

\begin{minipage}{13cm}

\vspace*{12pt}

{\begin{small}
Fig.~1. Comparison of our model with the preliminary 
CERES data \cite{CERES_Pb} (upper panels, 
$Q_\perp$ spectrum for $M = 0.25 \cdots 0.68$ GeV) and 
NA50 data \cite{NA50} (lower panels,
$Q_\perp$ spectrum for $M = 1.5 \cdots 2.5$ GeV).
The solid curves are the thermal yields, and the uppermost curves
depict the sum of all contributions.
For further details see text.
\end{small}}
\end{minipage}
\end{center}

\subsection{Sulfur beam data}

Let us now turn to the older sulfur beam data (cf.\ \cite{RappWambach}
for a recent survey). Since a much larger rapidity interval is covered
(see fig.~2)
we smear the source distribution (1) by a Gaussian function with a width
of 0.8. For $Y_{cms}$ we choose 2.45 
as suggested by an analysis of the hadronic
rapidity distribution (cf.\ fig.~3a).
For the CERES $e^+ e^-$ data \cite{CERES_S} in S(200 AGeV) + Au reactions,
$\frac{d N_{ch}}{d \eta} = 125$ and the published hadronic cocktail is used.
The NA38 $\mu^+ \mu^-$ data \cite{NA38} in the Drell-Yan region
$M > 4.2$ GeV can be used to pin down the intrinsic transverse momentum 
distribution
of partons. A value of $\langle k_\perp^2 \rangle \approx (0.8 \cdots 1)^2$
GeV${}^2$ is found (see fig.~3b), which we use in all PYTHIA simulations.

\begin{center}
\centering
~\\[-.8cm]
\hspace*{6mm}
\psfig{file=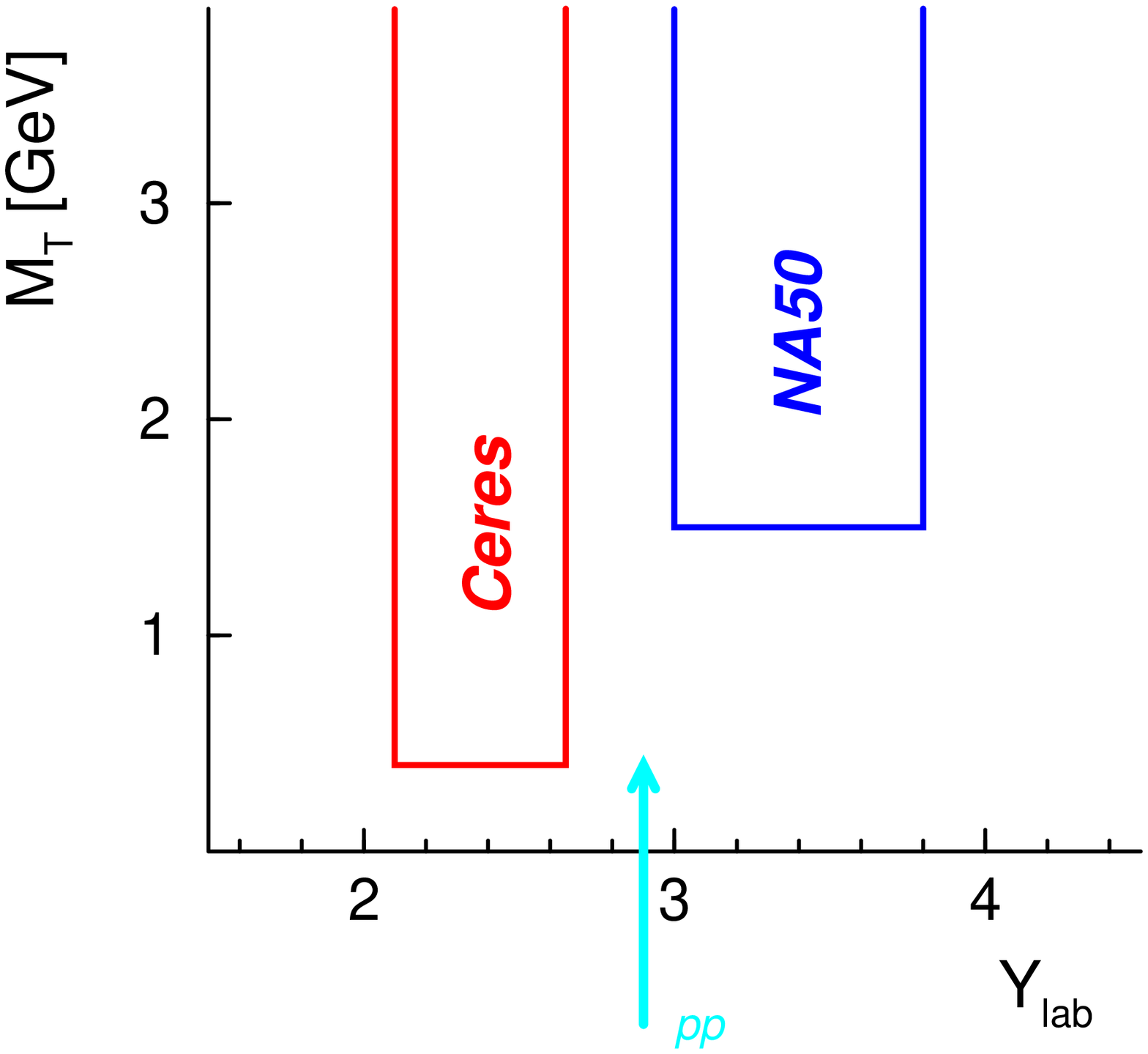,width=4.9cm,angle=-90}
\hfill
\psfig{file=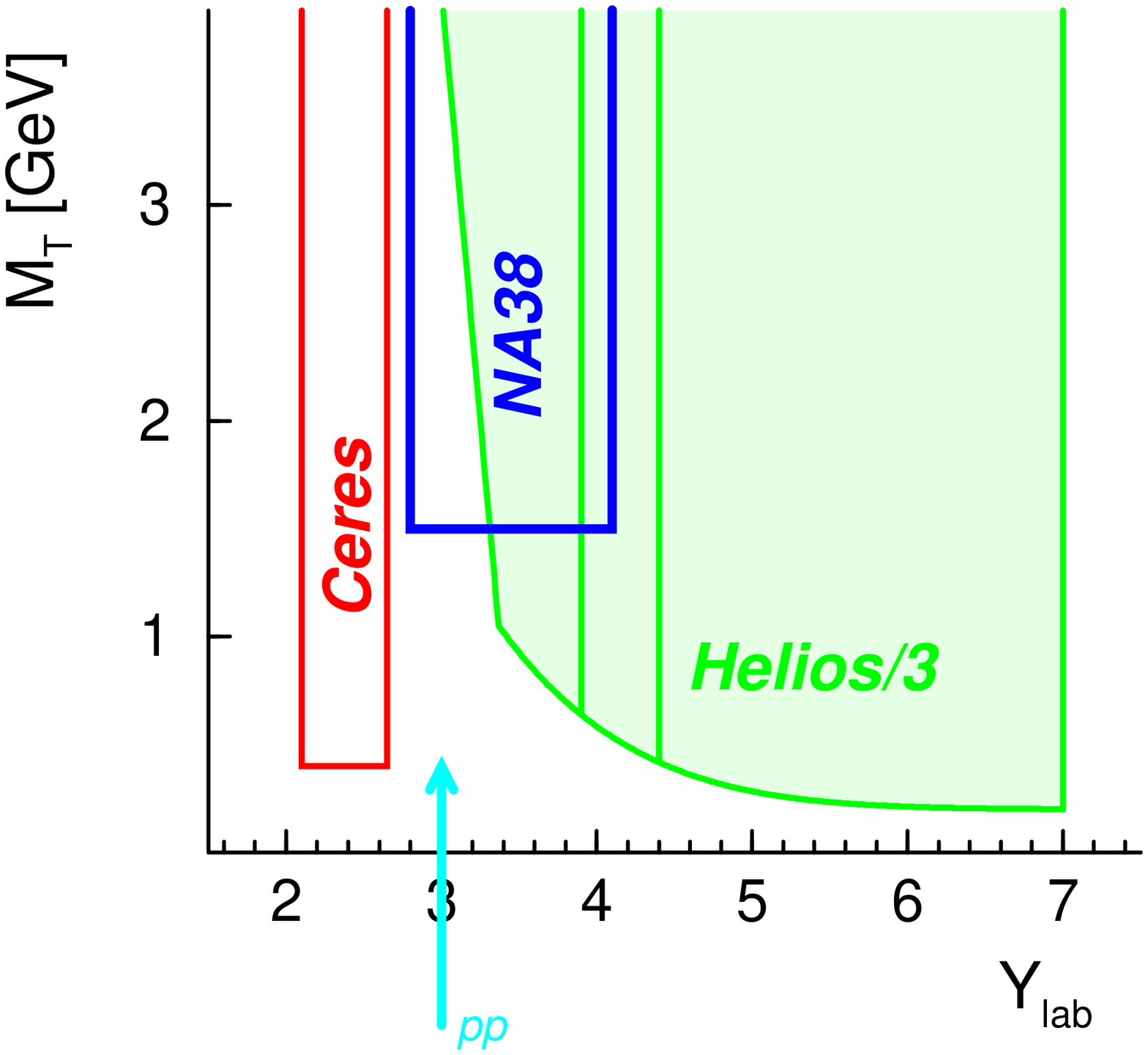,width=4.9cm,angle=-90}
\hspace*{6mm}

\begin{minipage}{13cm}

\vspace*{12pt}

{\begin{small}
Fig.~2. 
Coverage of the rapidity $Y_{lab}$ and transverse mass $M_\perp$ of the
various dilepton experiments.
(a) left panel: lead beam;
(b) right panel: sulfur beam.
\end{small}}
\end{minipage}
\end{center}
\begin{center}
\centering
~\\[-.6cm]
\hspace*{6mm}
\psfig{file=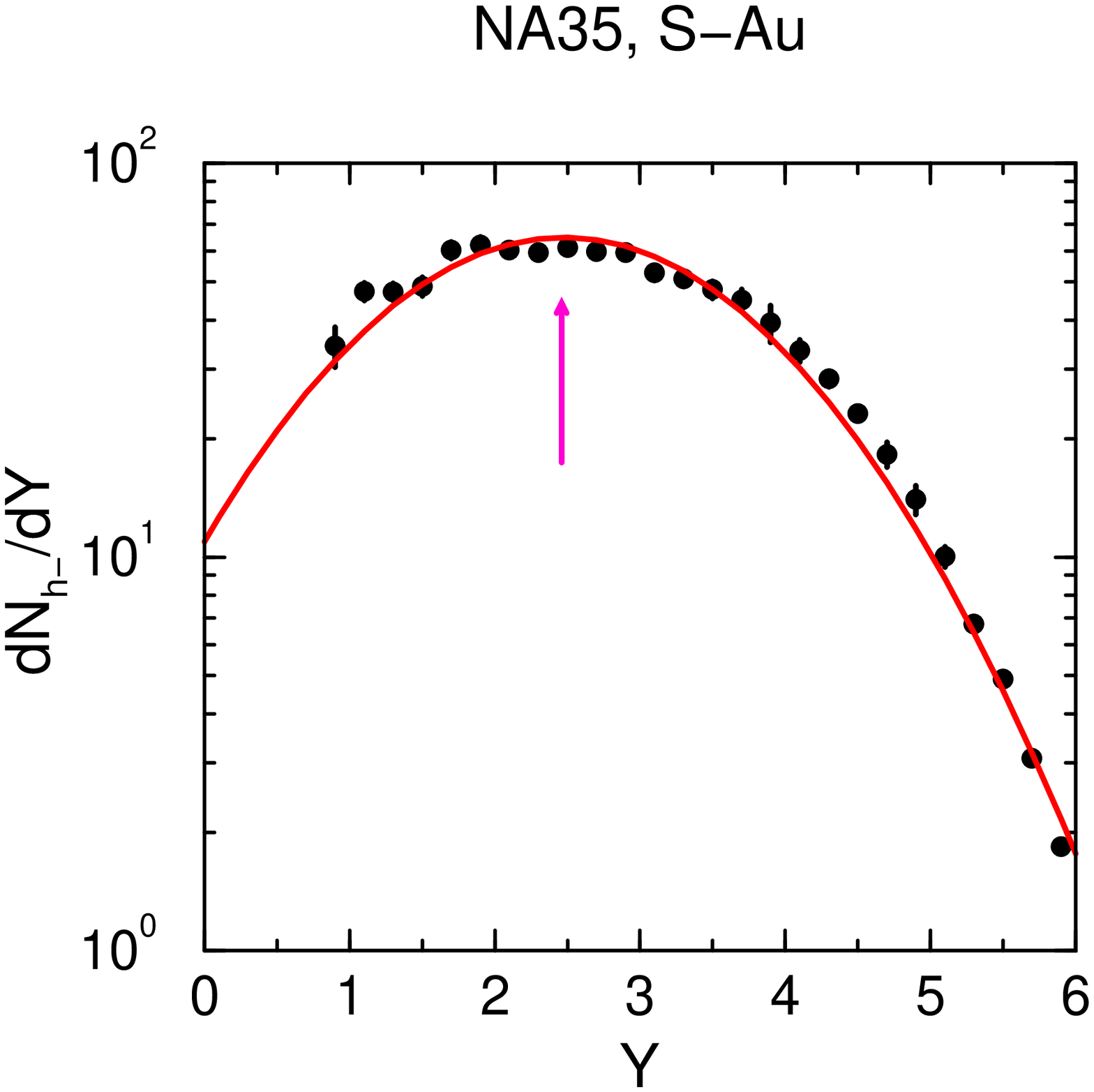,width=5.3cm,angle=-90}
\hfill
\psfig{file=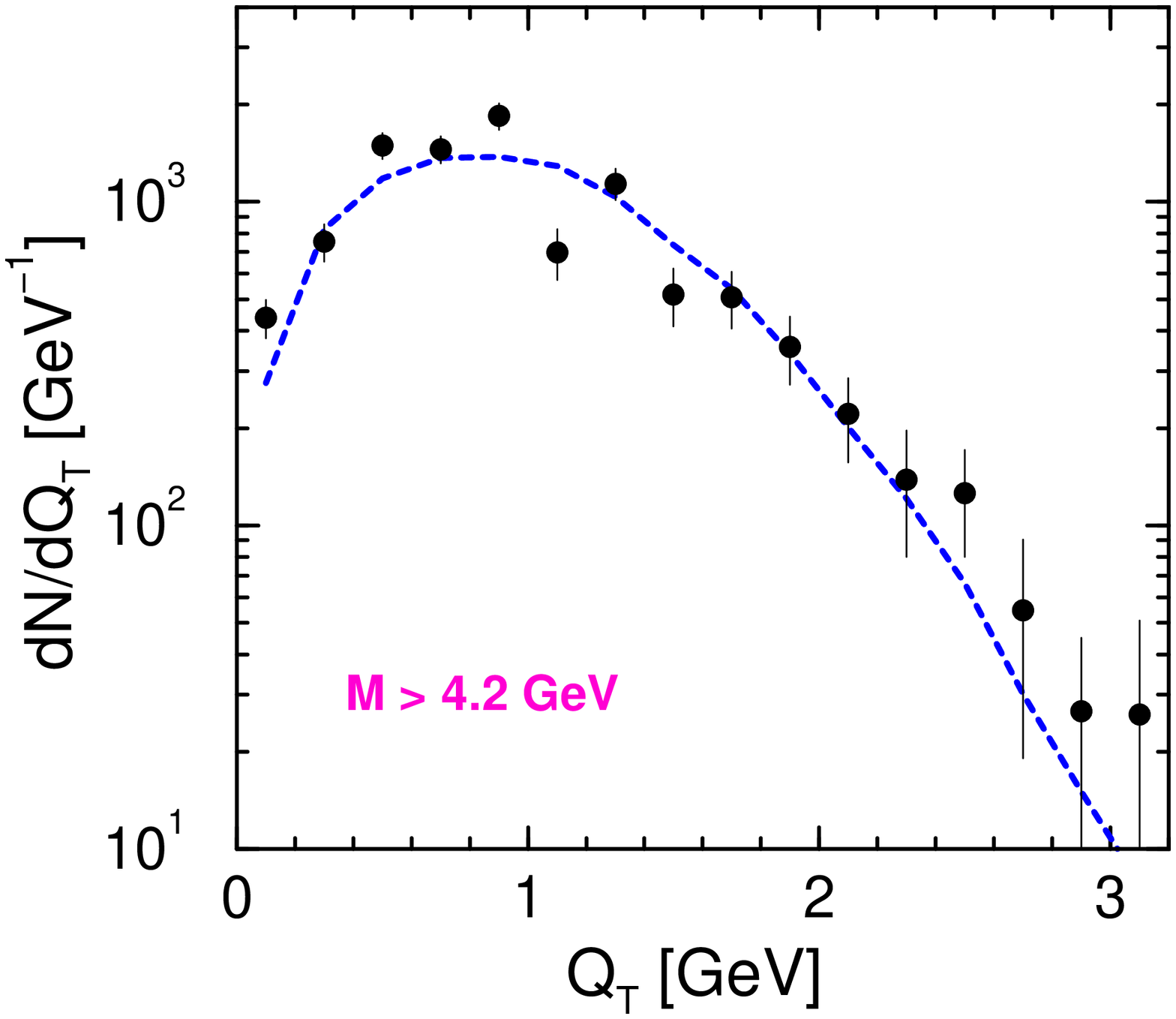,width=5.3cm,angle=-90}
\hspace*{6mm}

\begin{minipage}{13cm}

\vspace*{12pt}

{\begin{small}
Fig.~3. 
(a) left panel: Rapidity distribution of negatively charged hadrons 
(from \cite{NA35})
and a Gaussian fit centered at $y = 2.45$;
(b) right panel: Transverse momentum distribution of dileptons in the Drell-Yan
region in the NA38 experiment \cite{NA38}.
\end{small}}
\end{minipage}
\end{center}

\begin{center}
\centering
~\\[-.8cm]
\hspace*{6mm}
\psfig{file=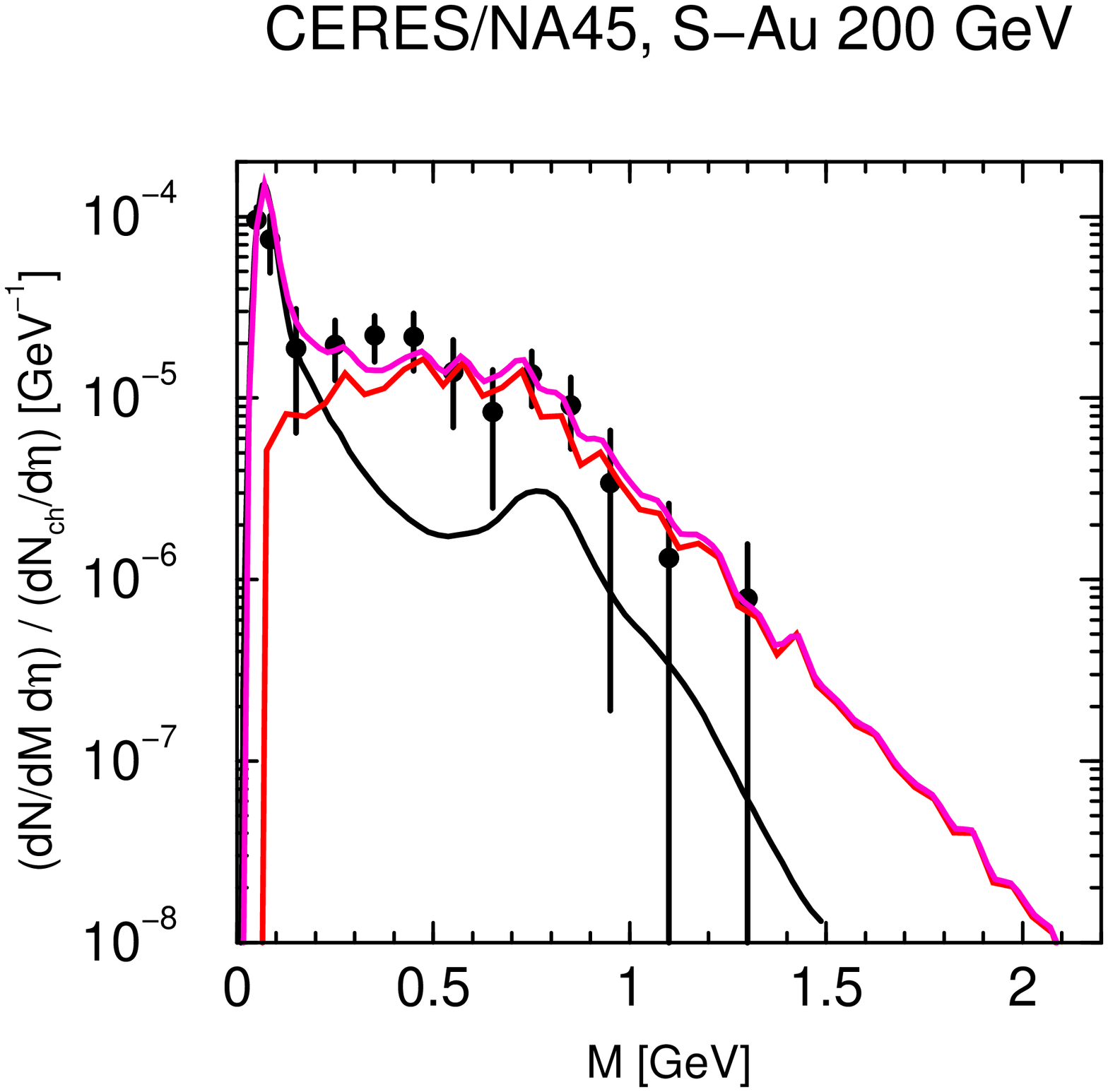,width=4.9cm,angle=-90}
\hfill
\psfig{file=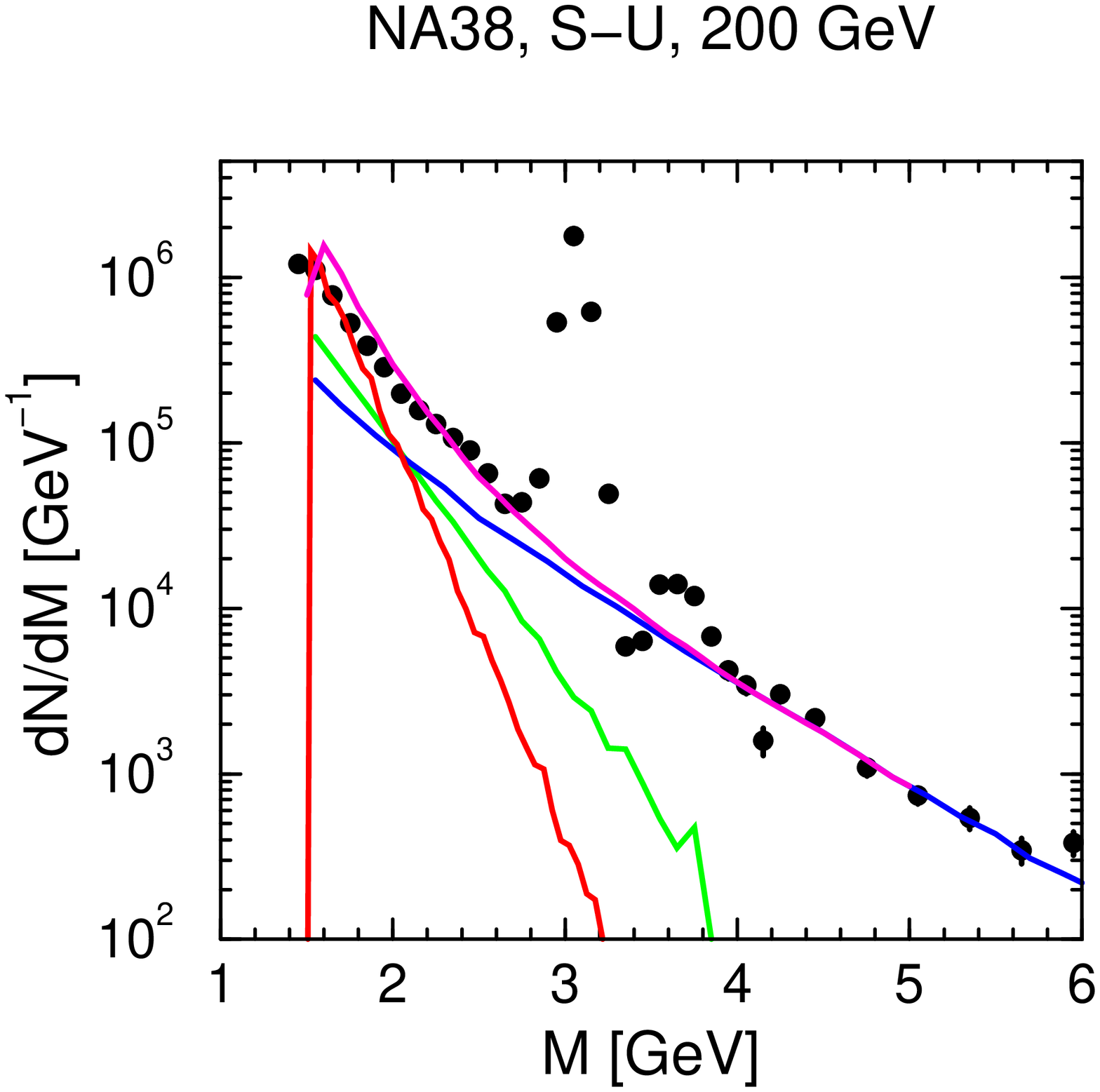,width=4.9cm,angle=-90}
\hspace*{6mm}

\psfig{file=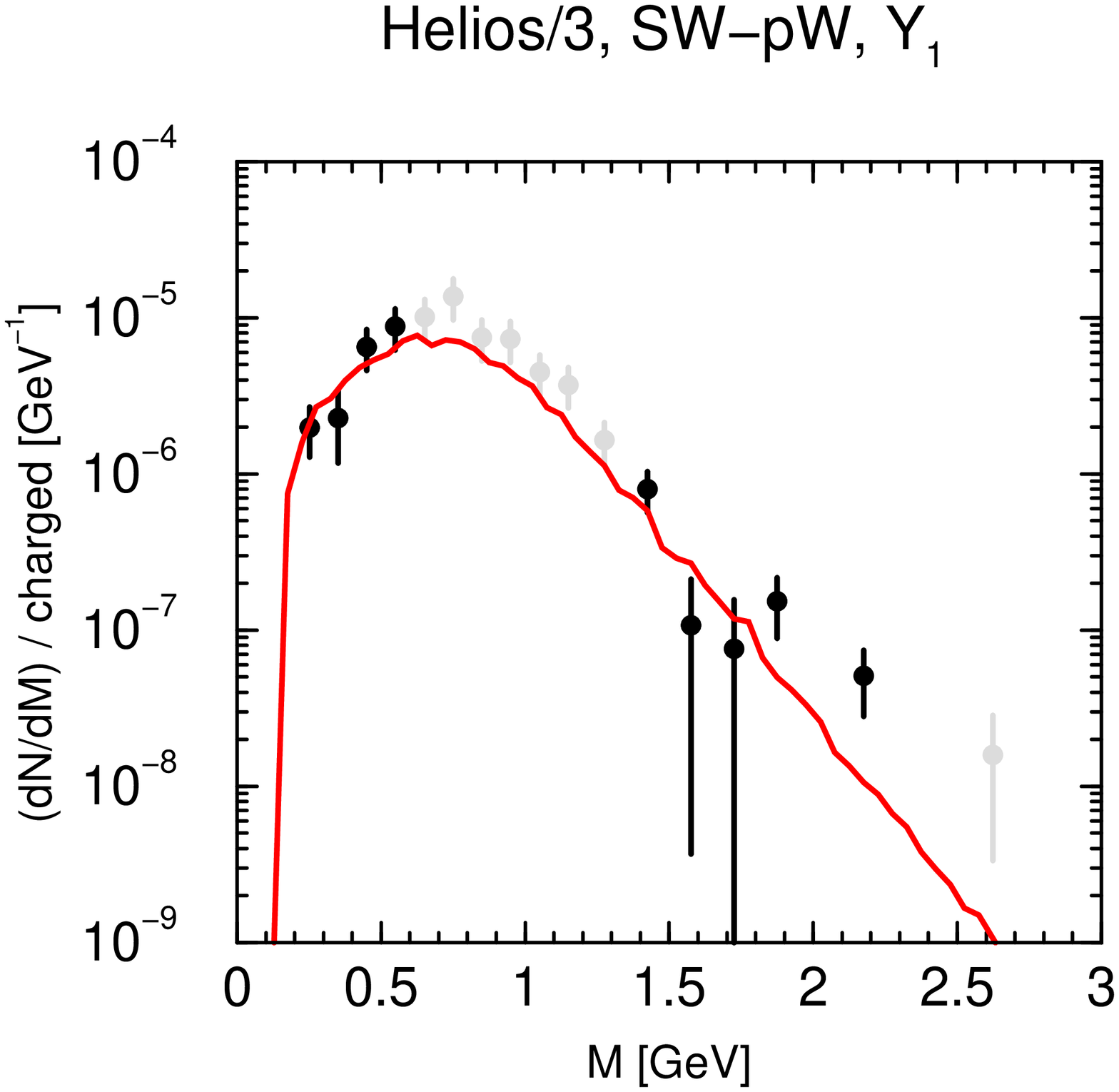,width=4.4cm,angle=-90}
\hfill
\psfig{file=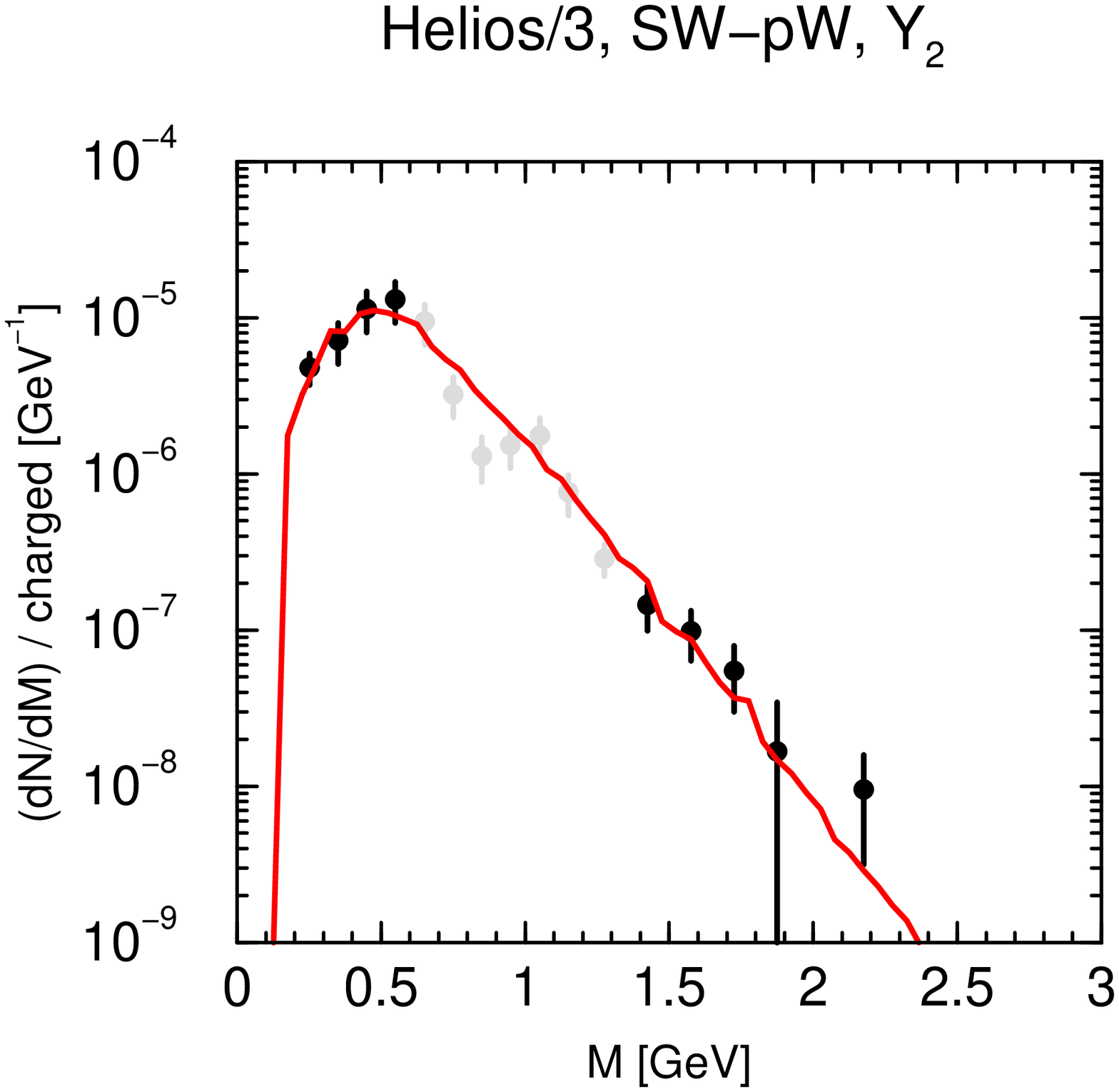,width=4.4cm,angle=-90}
\hfill
\psfig{file=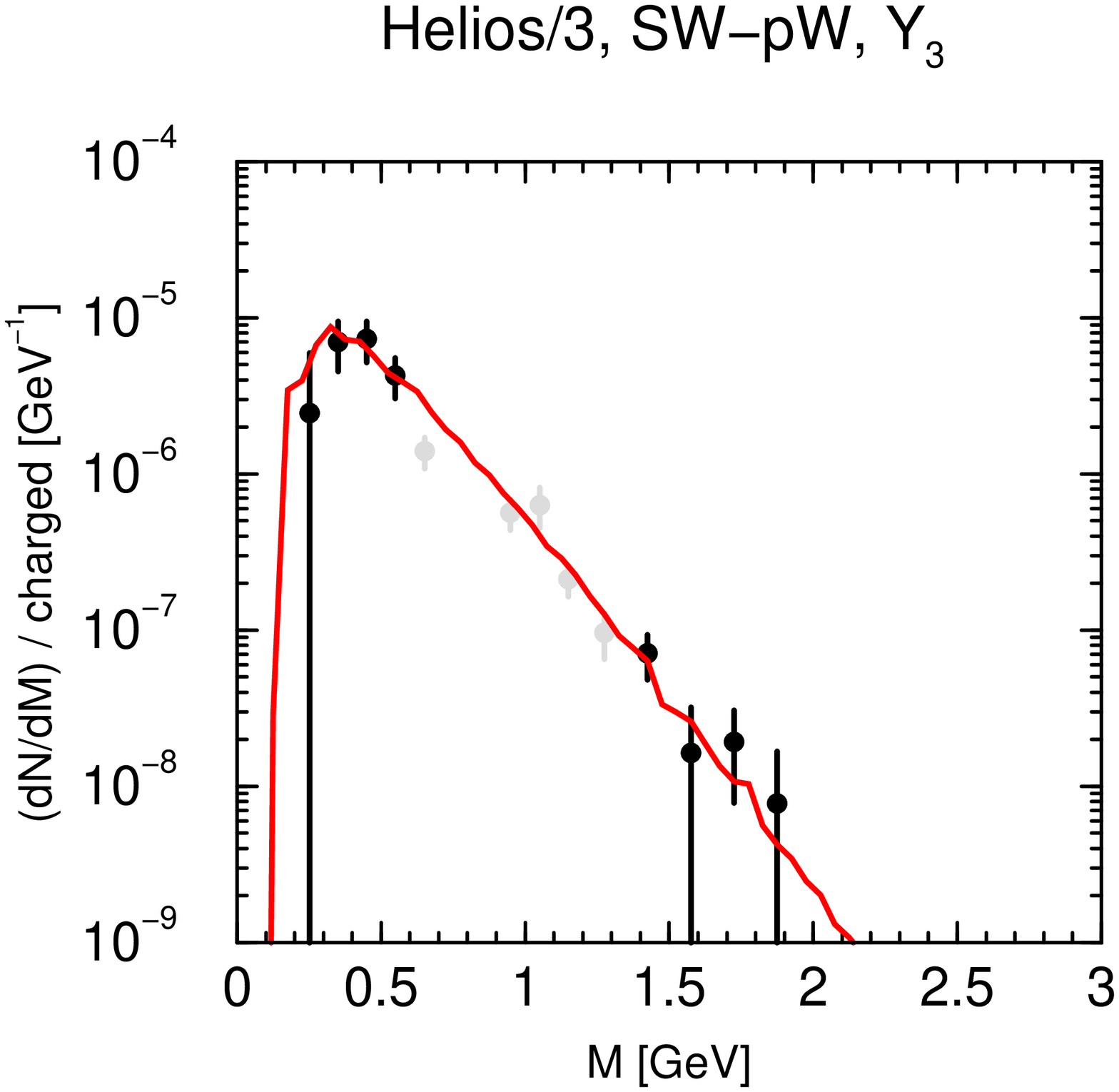,width=4.4cm,angle=-90}

\begin{minipage}{13cm}

\vspace*{12pt}

{\begin{small}
Fig.~4. 
Comparison of our model calculations with the data
\cite{CERES_S,NA38,HELIOS-3} with separately adjusted normalization factors.
Meaning of the curves: 
(a) upper left panel: lower solid curve with $\rho$ hump: cocktail,
curve above cocktail: thermal yield;
(b) upper right panel: thermal yield, open charm contribution and Drell-Yan
(from left to right at larger $M$);
the uppermost curves in (a, b) display the sum of all contributions;
lower panels: solid lines: thermal yield.

\end{small}}
\end{minipage}
\end{center}

While we can nicely reproduce the Drell-Yan background for the HELIOS-3 experiment,
our PYTHIA simulations deliver another open charm contribution than the one used
in previous analyzes \cite{LiGale}. Since the accurate knowledge of the background
contributions is necessary prerequisite, we use therefore for our analysis the
difference $\mu^+ \mu^-$ spectra of S(200 AGeV) + W and p(200 GeV) + W 
reactions 
\cite{HELIOS-3},
thus hoping to get rid of the background since these 
spectra are appropriately normalized.   

The comparison of our calculations with the data of CERES \cite{CERES_S} and
NA38 \cite{NA38} and HELIOS-3 \cite{HELIOS-3} are displayed in fig.~4
for $T_{\rm eff} = 160$ MeV. One observes a fairly well reproduction of the
spectral shapes.
The available transverse momentum spectrum of NA38 is also nicely reproduced
in shape (see fig.~5a). 
We leave here the normalization $N_{\rm eff}$ as free parameter to achieve an optimum
description of the data. It turns out that the found normalization factors for the
different data sets are quite different. 
We therefore conclude that a unique description
of the sulfur beam data within statistical errors
is not possible even though the target nuclei have very
similar masses.

We mention that adjusting the normalization to the CERES data
\cite{CERES_S} 
the published upper bounds of the direct photon yields \cite{WA80} 
are partially below our
model calculations when adopting the model described in section~2 
for photons too (see fig.~5b).

\begin{center}
\centering
~\\[-.4cm]
\hspace*{6mm}
\psfig{file= 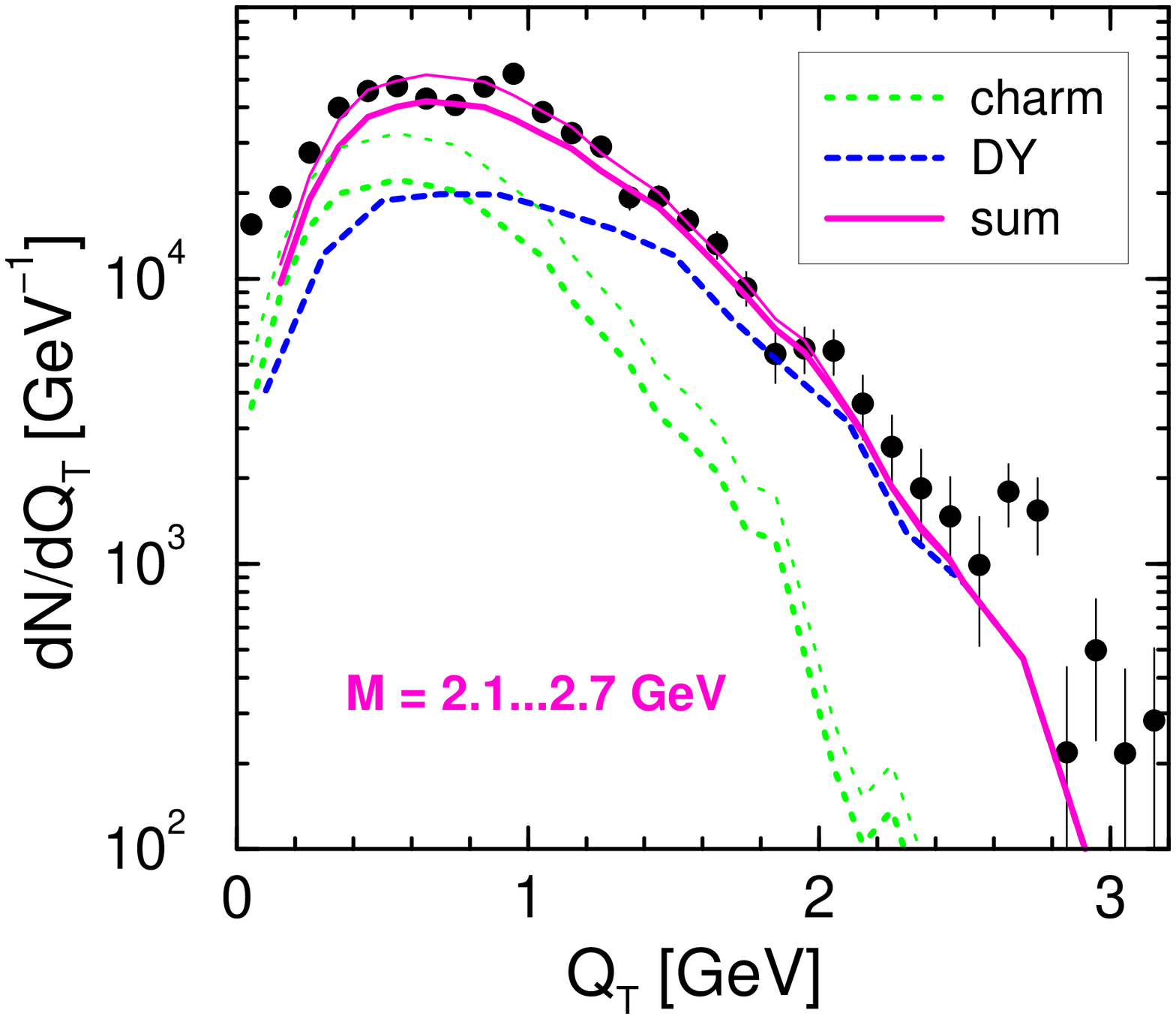,width=5.6cm,angle=-90}
\hfill
\psfig{file=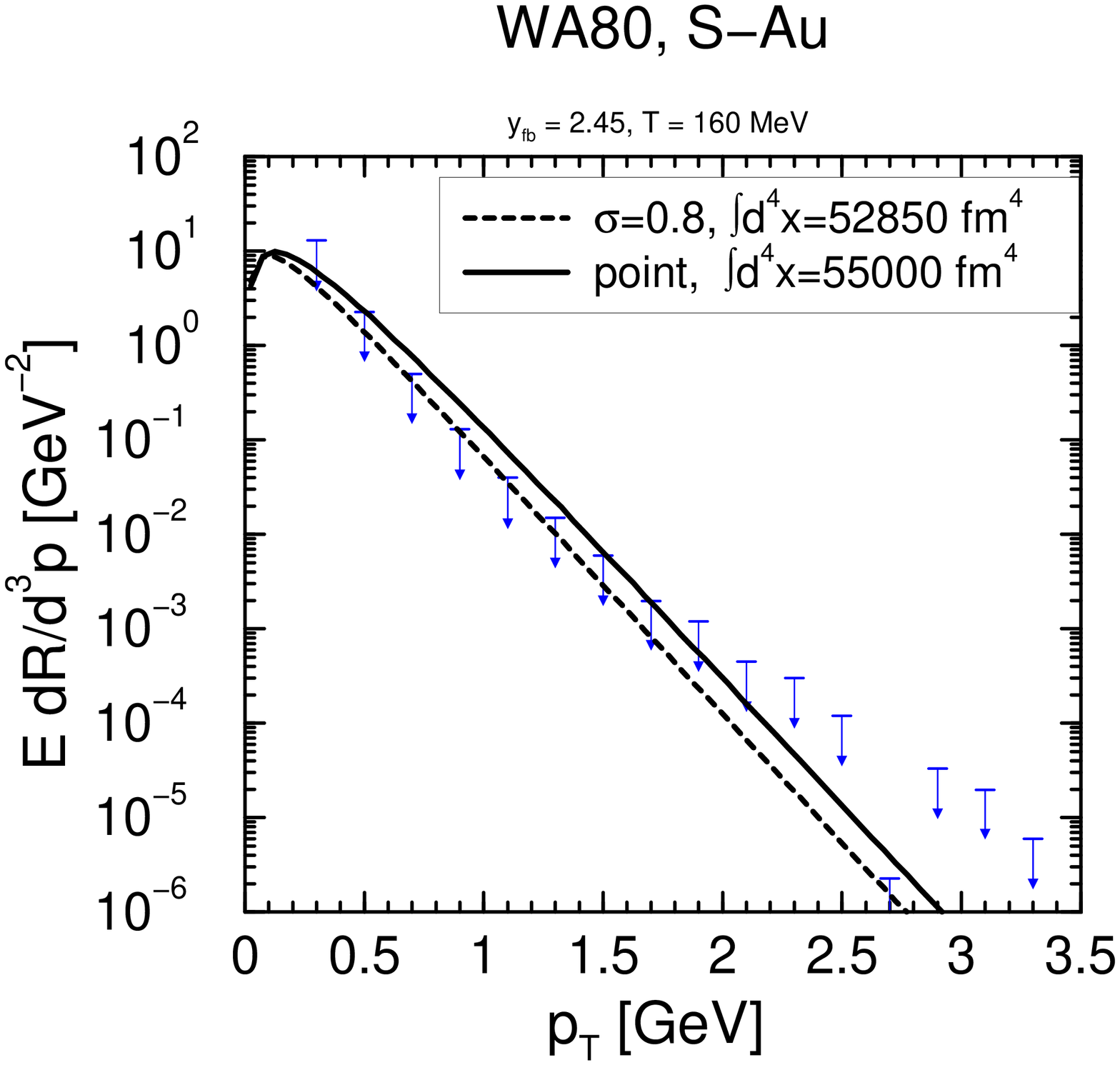,width=5.6cm,angle=-90}
\hspace*{6mm}

\begin{minipage}{13cm}

\vspace*{12pt}

{\begin{small}
Fig.~5. 
(a) left panel:
Comparison of our model calculations with the transverse momentum spectrum of
dileptons from NA38 \cite{NA38} in the intermediate-mass region.
(b) right panel: A comparison of the photon spectrum with the experimental
upper bounds \cite{WA80} when adjusting the source strength to the CERES data
\cite{CERES_S}; dashed curve: Gaussian smearing of the source,
solid curve: point like source.
\end{small}}
\end{minipage}
\end{center}

\section{Summary}

An attempt is reported to explain the dilepton data in recent CERN-SPS
experiments with heavy-ion beams. In doing so we employ the duality hypothesis,
i.e.\ we assume that due to strong in-medium effects the hadron spectrum is
drastically changed and the resulting dilepton emissivity looks as the one
of $q \bar q$ annihilation. The applicability of this hypothesis
relies on the
chiral symmetry restoration expected at temperatures achieved in the given
beam energy range.
While we find a good overall reproduction of the shapes of the experimental
spectra, only the lead beam data can be explained with a unique and reasonable
normalization.

For next future an analysis of the $E_\perp$ dependence of the NA50 data is envisaged.   

\end{document}